\documentclass[%
 reprint,
 amsmath,amssymb,
 aps,
floatfix,
]{revtex4-2}

\usepackage{graphicx}
\usepackage[caption=false]{subfig} 

\usepackage{dcolumn}
\usepackage{bm}
\usepackage{verbatim} 
\usepackage{hyperref}
\hypersetup{
colorlinks=true,
linkcolor=blue,         
citecolor=blue,        
filecolor=blue,        
urlcolor=blue
}

\begin{document}

\preprint{APS/123-QED}

\graphicspath{ {./figures/} } 

\title{Compact active vibration isolation and tilt stabilization\\
 for a portable high-precision atomic gravimeter}

\author{F. E. Oon\textsuperscript{1,}}
\email{fong\_en@nus.edu.sg}
\author{Rainer Dumke \textsuperscript{1,2}}%

\affiliation{%
 \textsuperscript{1} Centre for Quantum Technologies, National University of Singapore, 3 Science Drive 2, 117543, Singapore
 \\
 \textsuperscript{2} Division of Physics and Applied Physics, Nanyang Technological University, 21 Nanyang Link, 637371, Singapore 
}%

\date{\today}

\begin{abstract}
In high-precision atomic gravimeters, a rest mass is needed to provide a gravity reference, which is typically the retro-reflecting mirror of the Raman laser beams that addresses the two hyperfine ground states of the specific free-falling atomic test mass. We constructed a compact active feedback control system for the retro-reflecting mirror that provides vibration isolation in the vertical axis as well as two rotation stabilization in the horizontal plane. The active feedback control provides vertical vibration reduction of up to a factor of 300 in the frequency range of 0.03 to 10 Hz and tilt stabilization of approximately $\pm$ 1 \textmu{}rad in the noisy lab environment. This system has enabled high-precision gravity measurements on a portable gravimeter with sensitivity reaching $6.4\times 10^{-8}$ g/$\sqrt{\text{Hz}}$ and resolution of $2.8\times 10^{-9}$ g after an integration time of 4000 s. 

\end{abstract}

\maketitle

\section{
\label{sec:level1}
MOTIVATION}

In recent years, enormous efforts have been put into transferring laboratory-based quantum gravimeters into field-applicable devices~\cite{menoret2018gravity,chen2020portable,bidel2018absolute}. The main advantages of a free-fall atomic gravimeter are that it does not suffer from bias drift and provides absolute gravity measurements. Transportable atomic gravimeters open numerous applications such as monitoring magma buildup in volcanoes, ice mass changes, exploring underground minerals or geothermal resources, as well as applications in inertial navigation ~\cite{bongs2019taking, van2016slightness, stray2022quantum,cheiney2018navigation}.    
\par

The major challenge of field measurements in a noisy environment is vibration isolation at low frequencies~\cite{hensley1999active}. For high-precision gravimeters based on two-photon stimulated Raman transitions~\cite{peters2001high}, the sensitivity of gravitational detection scales quadratically with the time separation between the interferometry pulses, $T$, which is usually in the range of 100$-$200 ms. The interferometric measurement is less sensitive to high vibration frequencies, as the gravity measurement behaves like a second-order low-pass filter with corner frequency of $1/T$~\cite{cheinet2008measurement}. Hence, to resolve the gravitational variations at the few \textmu Gal level, the vibrations at frequencies less than a few hertz have to be subjugated at least at the same level.

\par
In order to reduce the physical size of the vibration isolation system, a commercially available passive vibration isolation stage can be used (see Feier et al.~\cite{ChristianThesis}), with a resonance frequency of approximately $0.5$ Hz. To further suppress the low-frequency vibrations, the acceleration of the stage is measured by an accelerometer and then actively fedback to a voice coil actuator that provides counterbalancing forces to the passive vibration isolation stage, effectively decreasing the resonance frequency to approximately $0.03$ Hz; however, the passive isolation stage has to be  modified to accommodate the actuator, and there exist no means to stabilize the drifts of the tilt angles of the passive isolation stage. A few improvements have since been made~\cite{zhou2012performance, tang2014programmable, zhou2020active}, however the tilt control of the stage is still lacking, and the physical dimensions of the vibration isolation system are still a major obstacle in realizing a compact absolute atomic gravimeter. 

\begin{figure*}
\includegraphics[width=1\textwidth]{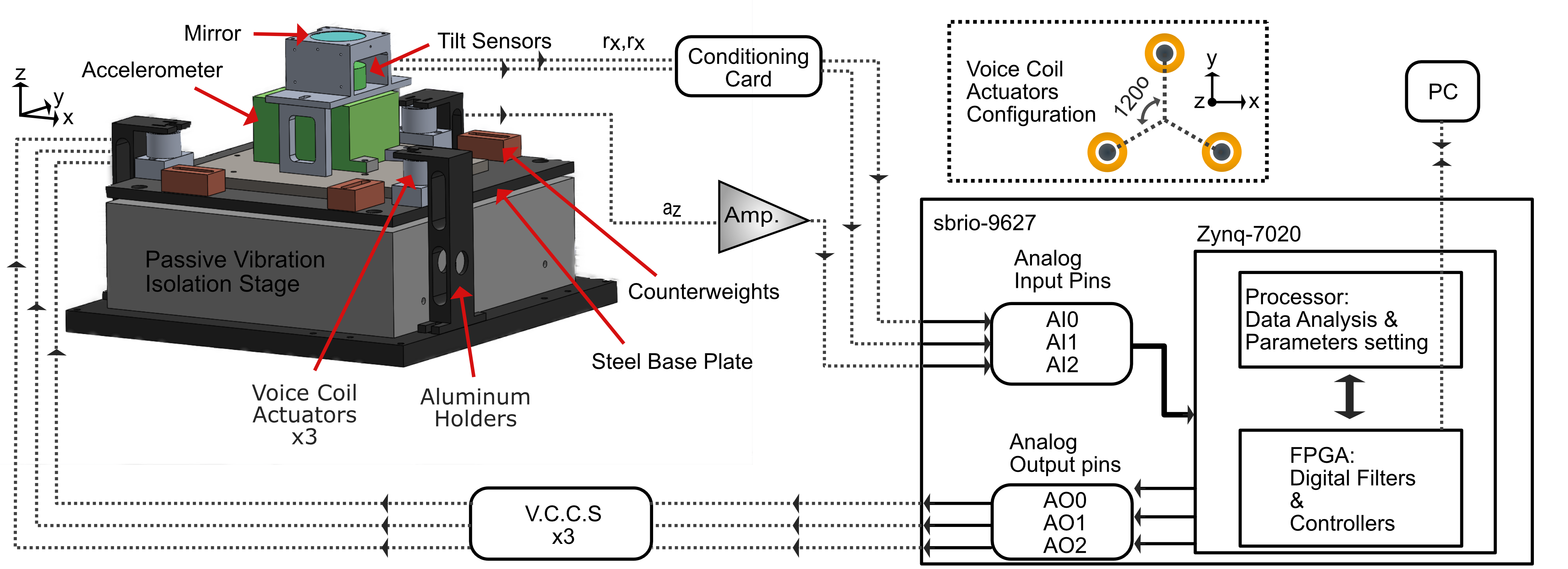} 
\caption{\label{fig:ActiveCancellationSchematic} The retro-reflecting mirror, the high-precision dual axial tilt sensors and the accelerometer are located at the center of the passive isolation stage. The signal of the accelerometer is first amplified by 19 times and offset to about 0 V then followed by another amplification of 10 times before being sampled by the analog input of the data acquisition device (National Instruments, sbrio-9627), while the signals from the tilt sensors go through a condition card (Jewell Instruments, model 83162) before being sampled by another two analog inputs. The filters for the signals and the controllers for the feedback forces are implemented digitally via an integrated field programmable gate arrays (FPGA) and real-time signal analysis (such as Fourier spectrum of the input signals) is performed via the built-in processor. The current of each voice coil actuator is provided individually by three home-built voltage-controlled current sources (VCCS). Each input and output sequence takes 10 \textmu s (10-kHz sampling rate). Components not shown: quarter-wave plate mounting right above the retro-reflecting mirror, thermoelectric cooler sandwiched between the accelerometer and the metal base, and a small Sorbothane cushion (tilt pad) sitting underneath the steel base plate.}  

\end{figure*}

In this work, we employ three voice coil actuators (BEI Kimco, LA13-11-001A) positioned at 120\textdegree{} with respect to each other and at equal distances from the center of the passive isolation stage (Minus K Technology, 25BM-10), as shown in Figure~\ref{fig:ActiveCancellationSchematic}. The feedback force in the vertical direction (z axis) is provided by applying equal current to the three actuators while the tilt adjustments (x and y axes) are provided by applying imbalance current in the respective axis. This configuration avoids the need to modify the commercial passive isolation stage; it also lifts the restriction of placing the actuator in the center of the stage. To further reduce the physical size of the vibration isolation system, we adopt a compact triaxial force-balance accelerometer (Nanometrics, Titan). The physical dimensions of our active vibration isolation system are $365\times 365\times 280$ mm$^3$ and the total weight is approximately 27 kg. These represent at least approximately 40\% reduction in both weight and height compared to previous works~\cite{zhou2012performance, tang2014programmable, zhou2020active}. 


%

\section{IMPLEMENTATION}

\subsection{Experiment setup}


\begin{figure}
\includegraphics[width=0.5\textwidth]{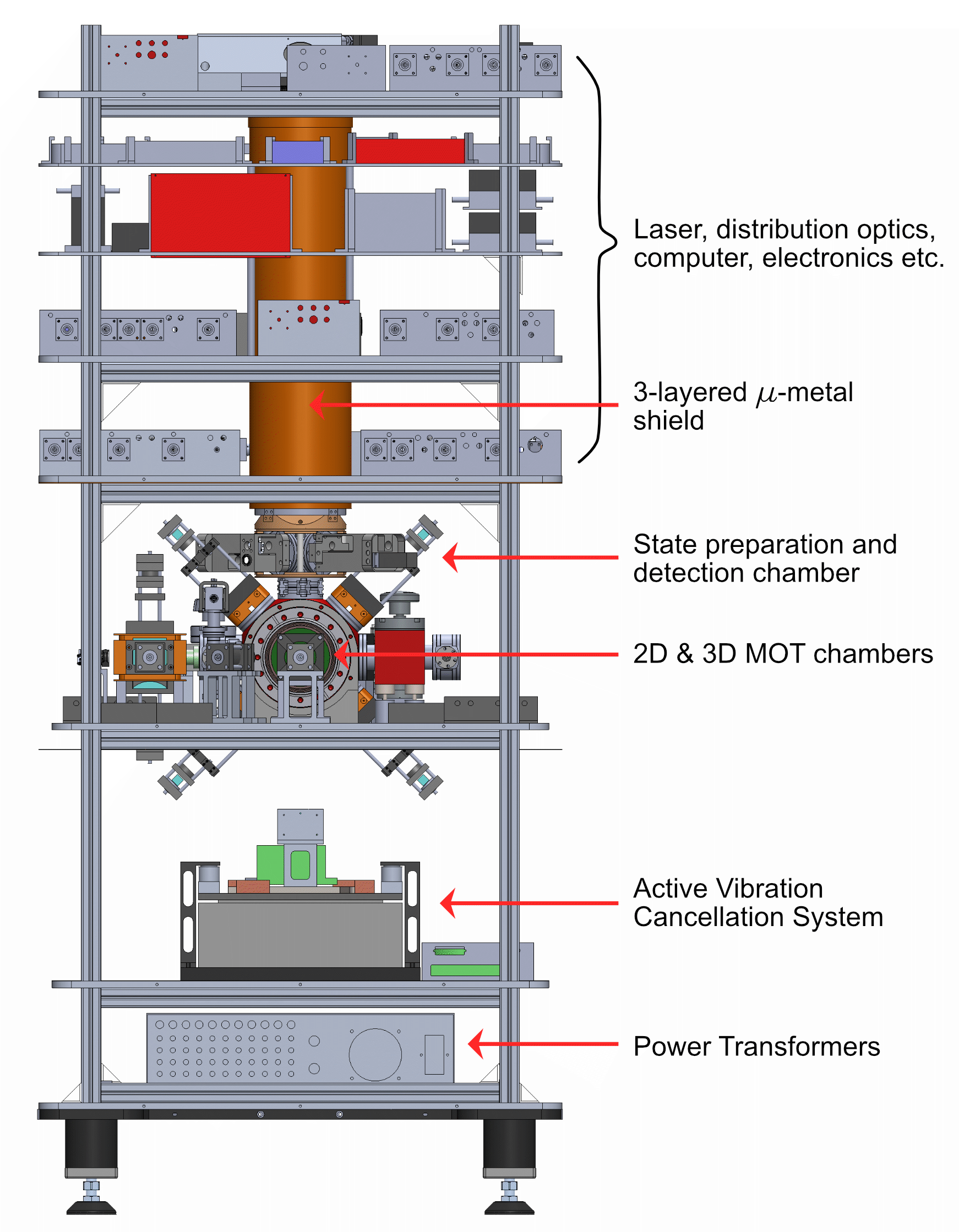} 
\caption[]{Computer-aided design illustration of the gravimeter. The vacuum chamber for the 3D MOT is located at the center of an ($800 \times  800 \times   1800$)-mm$^3$ movable structure. The customized optics and electronic components are located around the three-layered mu-metal shield. The retro-reflecting mirror and the active vibration cancellation system are located right underneath the 3D MOT chamber; hence the Raman laser pulses travel from top to bottom and are retro-reflected from the mirror. During routine measurements, the gravimeter is controlled remotely via an Internet connection.} 
\label{fig:ExperimentSchematic} 
\end{figure}

We employ $^{87}$Rb atoms as our gravity test mass. During each measurement cycle, the $^{87}$Rb atoms are first loaded and cooled via a combination of a two-dimensional (2D) magneto-optical trap (MOT) and three-dimensional (3D) MOT~\cite{en2021transportable,oon2022compact}. The ultracold atoms (approximately 3 \textmu K) are then launched vertically upward via the moving molasses technique~\cite{wynands2005atomic} where they undergo a free-fall parabolic trajectory. Right before entering a magnetic shielded cylindrical region (Figure~\ref{fig:ExperimentSchematic}), the atoms are prepared in a magnetic insensitive state of $F=2,\, m_F=0$ using two velocity-selection and blow-away sequences~\cite{peters2001high}. While inside the magnetic shielded region, the atoms are interrogated via Mach-Zehner matter wave interferometry, employing a Raman pulse sequence of $\pi/2$-$\pi$-$\pi/2$, with pulse separation time $T$, addressing the two hyperfine ground states $F=2,\, m_F=0$ and $F=1,\, m_F=0$ of $^{87}$Rb. Finally, after exiting the magnetic shield region, the phase difference acquired by the atoms and hence the value of local gravitational acceleration ($\Delta\phi=k_\text{eff}gT^2$, where $k_\text{eff}$ is the effective wave vector of the Raman laser pulses) are deduced via a state-selective detection. 

\par
The active vibration isolation system is located at the bottom of the gravimeter. To ensure a horizontal level, a high-precision two-axis tilt meter with resolution of 0.1 \textmu rad (Jewell Instruments, 755-1129) is used. The tilt meter and the accelerometer are placed right underneath the retro-reflecting mirror (see Figure~\ref{fig:ActiveCancellationSchematic}). Because of the large temperature drift of the accelerometer ($320\times 10^{-6}$g/\textdegree{}C), we temperature stabilize the accelerometer via a thermoelectric cooler to within $\pm 0.01$ \textdegree{}C. To provide a counterbalance force on the vertical axis and stabilization of the tilt angles of the passive isolation stage, we employ three voice coil actuators. The coil assemblies of the voice coil actuators are attached to three aluminum holders that are connected to the base of the passive isolation stage, while the magnets of the voice coil actuators are mounted on the platform of the passive isolation stage. The magnets of the voice coil actuators are physically disconnected from the coil assemblies to minimize the mechanical coupling of vibrations to the passive isolation stage. To provide force in the vertical z axis, we apply equal currents to the three identical voice coil actuators simultaneously. To compensate the rotation of the stage along the x axis, we independently control only the current of one actuator (sitting at the positive side of the y axis). To rotate the stage along y axis, we apply equal but opposite currents to the two actuators sitting at the positive and negative sides of the x axis. We can therefore stabilize the rotation along x and y axes independently.

\subsection{Digital controllers}

\begin{table}
\begin{center}
    \begin{tabular}{ | c | c | c | c | c | }
    \hline
 Acc. & Filters/controllers & K & $\omega_1/2\pi$ & $\omega_2/2\pi$
    \\ \hline
 & \textbf{HP} & 1 & 0.0008 & 
   \\ \hline
 &  \textbf{Lag1} & 1 & 600 & 1
   \\ \hline
 &  \textbf{Lag2} & 1 & 800 & 3
    \\ \hline
 & \textbf{Lag3} & 200 & 1 & 0.05
  \\ \hline
 & \textbf{Lead1} & 20 & 10 & 100
  \\ \hline
 Tilt. & & & & 
 \\ \hline
 &  \textbf{Lag4} & 1 & 1000 & 50
  \\ \hline
   &  \textbf{Lag5} & 500 & 0.1 & 0.0001
  \\ \hline
 
    \end{tabular}
\end{center}
\caption[]{Digital controllers for vibration cancellation in the vertical axis are $\text{D(s)}=\text{HP}^2\times \text{Lag1}\times \text{Lag2}\times (\text{Lag3}+\text{Lead1})$ while the digital controllers for tilt stabilization in both x and y axes are $\text{D(s)}=\text{Lag4}\times \text{Lag5}$. Two high-pass filters $\text{HP}$ eliminate the offset and drifts of low-frequency signals that are not due to actual vibrations. Lag compensators $\text{Lag1}$, $\text{Lag2}$ and $\text{Lag4}$ serve as the low-pass filters for the signals. $\text{Lag3}$ and $\text{Lag5}$ provide the stabilization force for the vibration and tilt stabilization respectively. $\text{Lead1}$ provides extra stabilization at high frequencies.}
\label{Table:DigitalFilters}

\end{table}

The subject of designing feedback controllers of dynamic systems can be found in Refs.~\cite{FeedbackControlGene, hensley1999active, en2021transportable}. The digital feedback controllers we adopt are in the form
\begin{equation}
\label{Eq:Dcontroller}
D(s)=K\frac{s/\omega_1+1}{s/\omega_2+1},
\end{equation}
\noindent with $s=i\omega$, where $\omega$ is the vibration frequency in rad/Hz and $K$ is the gain. We have lag compensator for corner frequency $\omega_1 > \omega_2$ and lead compensator for corner frequency $\omega_1 < \omega_2$. To eliminate the offset and drifts of the input signal from the accelerometer, we first condition the input signal via two high-pass filters in the form of

\begin{equation}
\label{Eq:HP}
HP=\frac{s/\omega_c}{s/\omega_c+1},
\end{equation}

\noindent where $\omega_c$ is the cutoff frequency. The exact filters and controllers implemented are shown in Table~\ref{Table:DigitalFilters}.

\section{Performance}

 \begin{figure}
\centering

\includegraphics[width=0.5\textwidth]{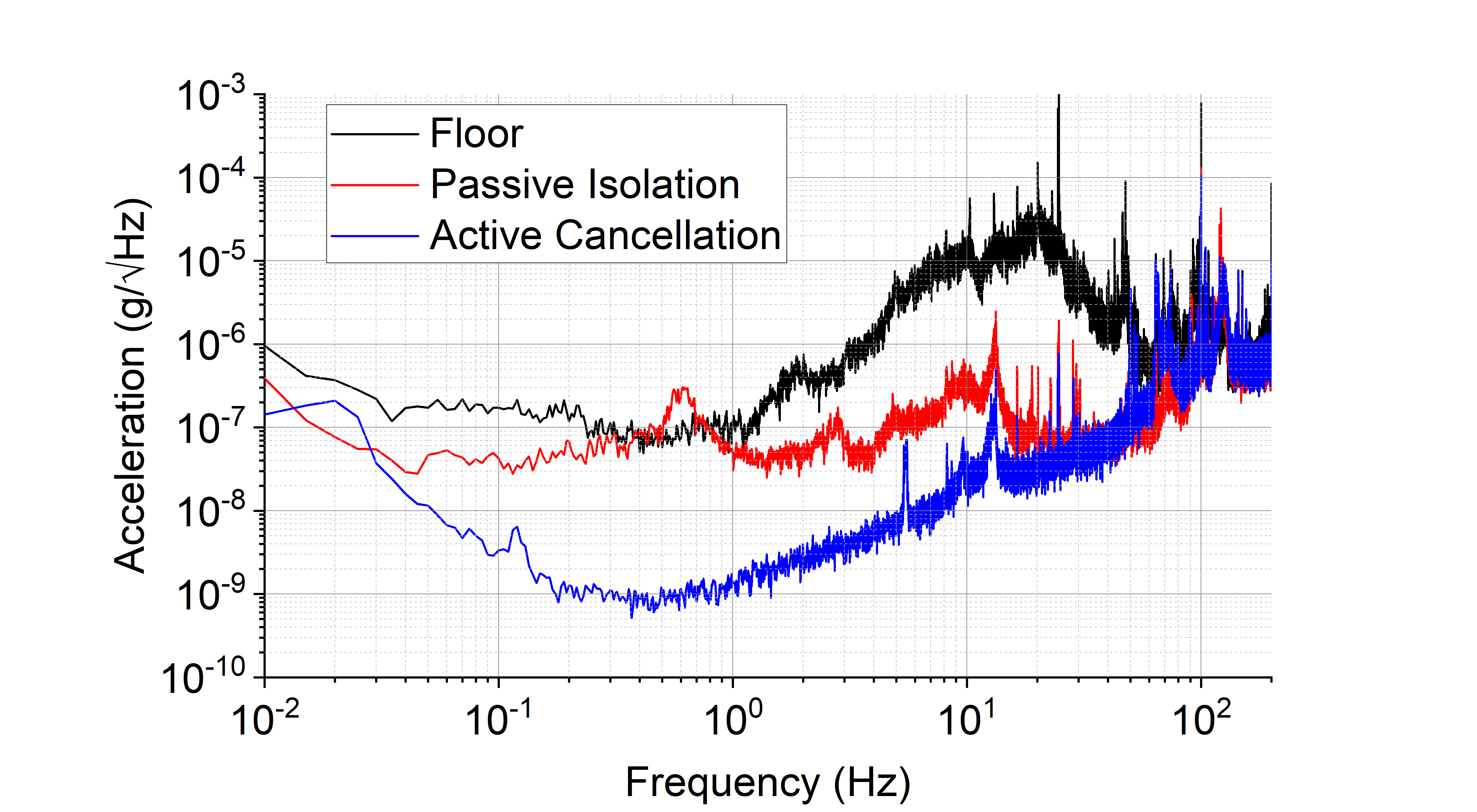} 
\caption[]{Black curve: residual vibrations of the laboratory which have a dominant vibration frequency at approximately 23 Hz. Red curve: vibrations frequencies at high frequencies are attenuated via the passive isolation stage. The resonant frequency of the passive isolation stage can be seen at $0.6$ Hz. Blue curve: residual vibrations of the retro-reflecting mirror with feedback loop activated. } 
\label{fig:ActivePerformance} 
\end{figure}

 \begin{figure}
\centering
\includegraphics[width=0.5\textwidth]{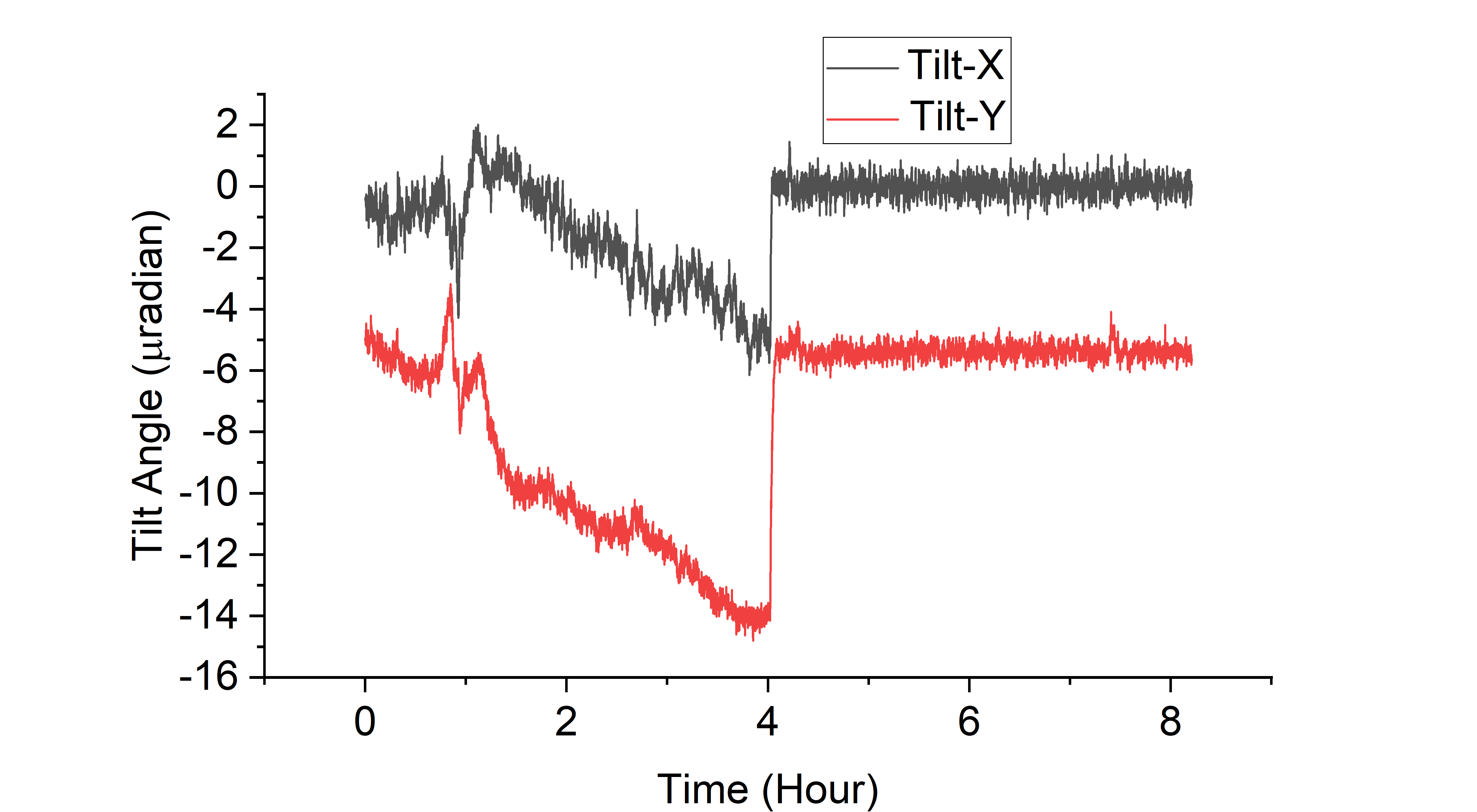} 
\caption[]{Rotation drifts of the passive isolation stage along x and y axes are recorded in the first 4 hours. Rotation stabilization is activated at the onset of the fourth hour, resulting in rotation stabilization of $\pm\,1$ \textmu{}rad. Poorer rotation stability is observed when vertical vibration stabilization is activated simultaneously with the rotation stabilization, resulting in long-term rotation oscillations of $\pm\,5$ \textmu{}rad. } 
\label{fig:xtilt} 
\end{figure}

Figure~\ref{fig:ActivePerformance} shows residual vibrations of the retro-reflecting mirror when sitting on the floor, on the passive isolation stage and with vibration cancellation activated. The $1/T$ low-pass characteristic of the gravimeter at $T=140$ ms dictates that the gravity measurement inherits $>100\times$ attenuation on vibration frequencies $> 10$ Hz~\cite{en2021transportable}. The further attenuation of the high vibration frequencies by the passive isolation stage makes sure that gravity measurements are not affected by high vibration frequencies. The active stabilization is able to attenuate the vibrations with attenuation of up to a few hundreds in the frequency range of $0.03-10$ Hz, reaching a magnitude of $10^{-9}$g. It is clear that the stabilization does not perform well for vibration frequencies $< 0.03$ Hz. This is due to the fact that the ``information'' provided by the accelerometer is diminishing at low frequencies which is dominantly caused by temperature drifts. Therefore, any strong vibration that occurs at such low frequencies would inevitably affect the gravity measurement.

Figure~\ref{fig:xtilt} shows the rotation drifts of the passive isolation stage (with 50-Hz low-pass filter). Without tilt compensation, the tilt angle $\theta$ of the mirror would drift more than 100 \textmu{}rad within 24 hours, resulting in a change of gravity measurement of $>\, 10^{-9}$g (factor of $|g|\cos\,\theta$). In order to compensate for the drifts in rotation, imbalanced forces are created along the x and y axes by the three voice coil actuators where the tilt pad (comes together with the commercial passive isolation stage) underneath the steel base plate is deformed to produce the required tilt compensation. Hence, this tilt compensation is a slow process. Replacement with a softer tilt pad would result in a faster tilt compensation, but it would also enhance the coupling between the tilt motions and the vertical acceleration. Performing active vibration cancellation in this manner will in fact deteriorate the noise performance of the gravity measurements. The initial calibration of the tilt meter (true vertical direction) is performed via looking for the tilt angle that produces the highest value of $g$, which is also the highest value of Raman frequency chirping rate at the center fringe (see Section~\ref{sec:gravityMeasurement}). In our experimental setup, a tilt compensation of 100 \textmu rad requires a current difference of about 15 mA between the three actuators, while for vertical vibration cancellation, a few milliamps of current is generally sufficient.

\section{Gravity Measurement}
\label{sec:gravityMeasurement}
 
 \begin{figure}
\centering
\includegraphics[width=0.5\textwidth]{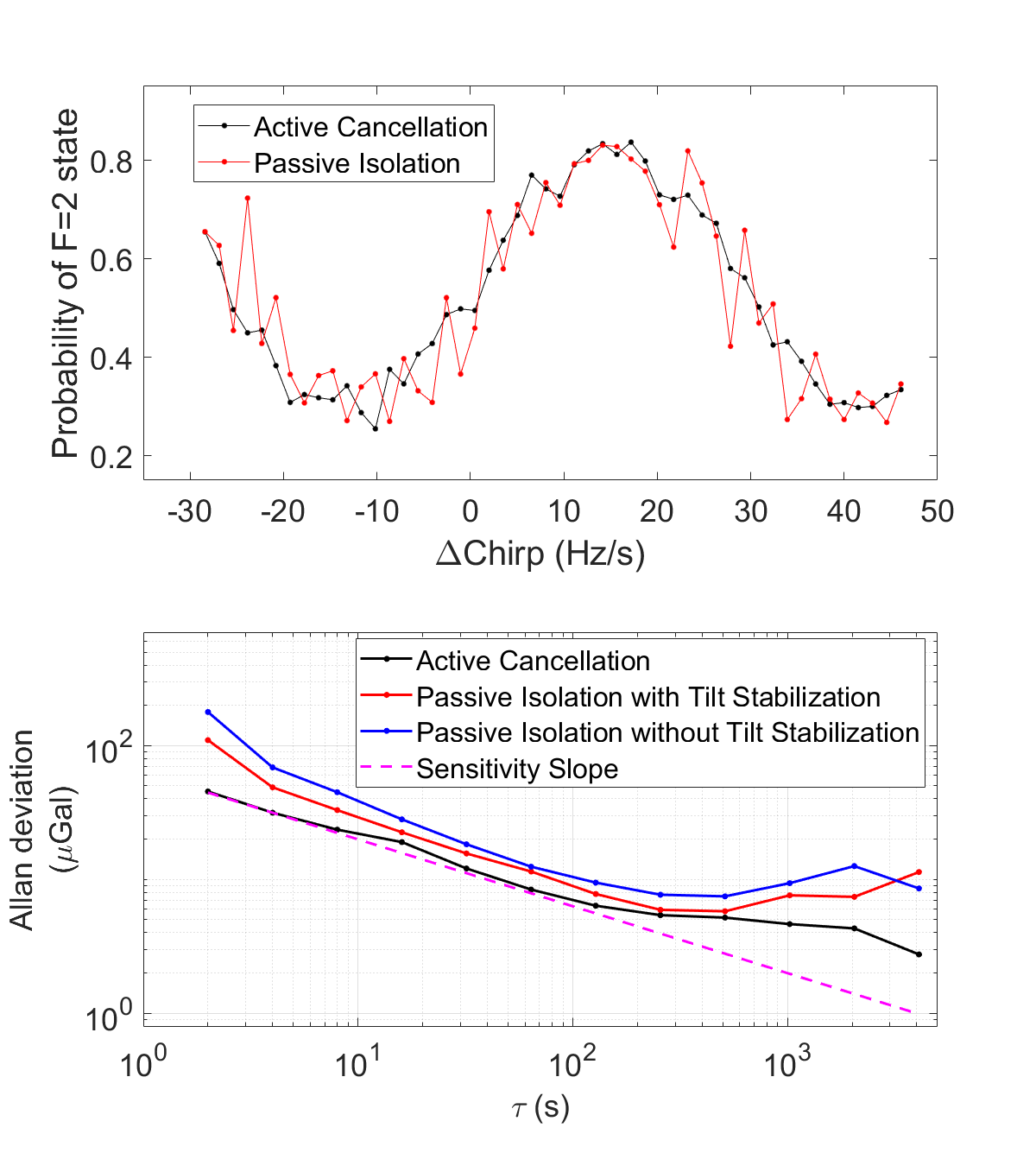} 
\caption[]{Top: interferometer fringes obtained by varying the frequency chirping rate of the Raman lasers. Each point represents a single measurement cycle, with a time interval of $2$ s. Red curve and black curve represent interferometer fringes at $T=140$ ms with passive vibration isolation and active vibration cancellation respectively. The fringe maxima represents the center fringe where the chirping of the Raman lasers cancels the gravity-induced phase shift of the atoms, $2\pi\Delta f T=k_\text{eff}gT^2$. Bottom: comparison of Allan deviations with and without active vibration cancellation. Dashed pink lines represent the gravimeter's sensitivity of approximately $63$ \textmu{}Gal/$\sqrt{\text{Hz}}$ ($6.4\times 10^{-8}$g/$\sqrt{\text{Hz}}$). We obtain a measurement resolution of $2.75$ \textmu{}Gal (approximately $2.8\times 10^{-9}$ g) after an integration time of $4000$ s. } 
\label{fig:fringe} 
\end{figure}

\begin{figure}

\includegraphics[width=0.5\textwidth]{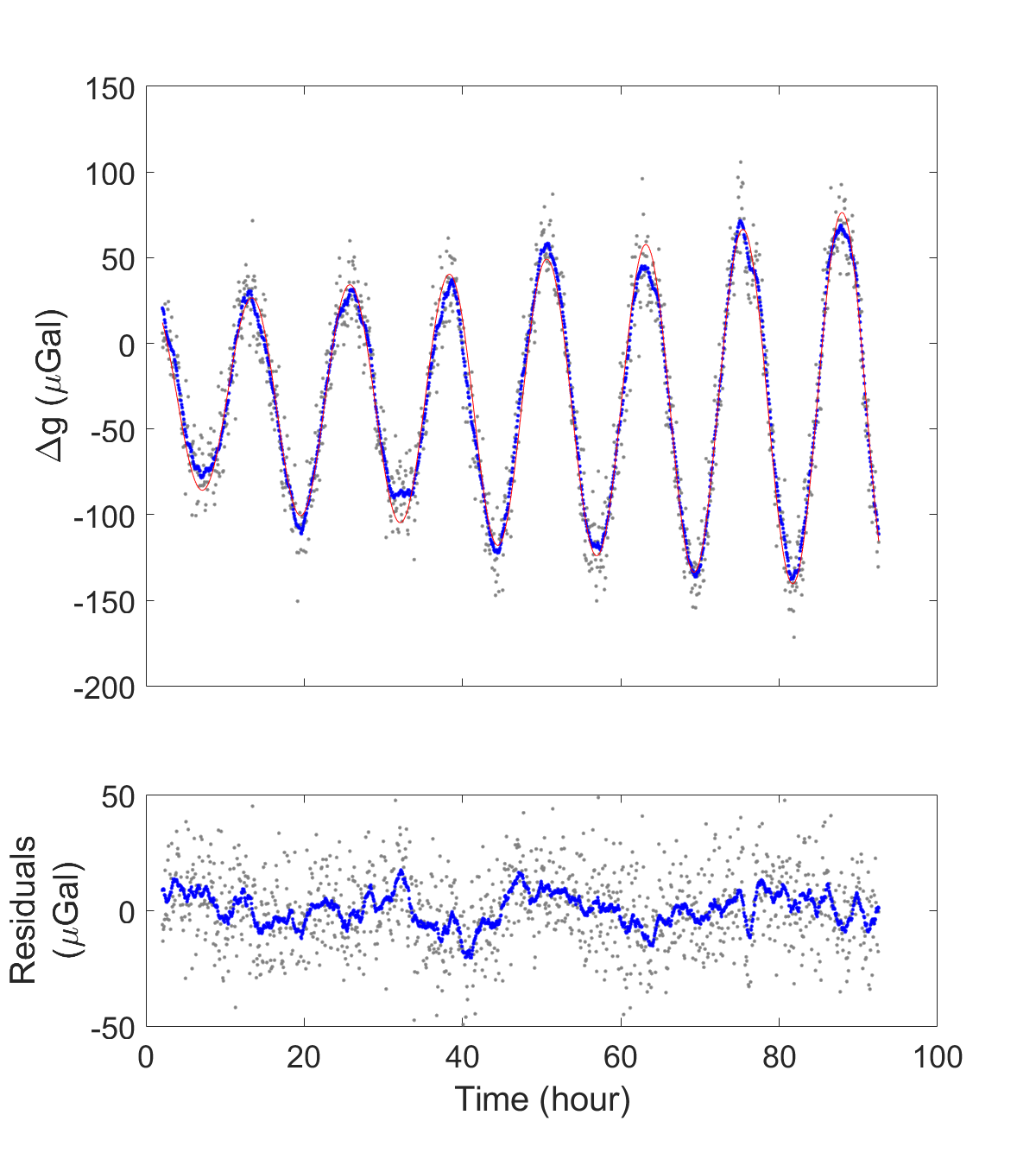} 
\caption[]{Uninterrupted gravity measurement performed at coordinates 1.34\textdegree{}N, 103.68\textdegree{}E, between 8th and 11st June, 2022. Red: tidal model extracted from Micro-G Lacoste, QuickTide Pro. Gray: each point represents a single center fringe scanning of a duration of 250 s (50 measurement sequences with 5-s measurement interval). Blue: 20-point moving average, equivalent to an integration time of approximately 83 minutes integration time. Bottom panel shows the residuals of gravity subtracted from Earth's tide model.} 
\label{fig:tidal} 
\end{figure}

During the interferometry sequence, the phase acquired by the free-fall atoms due to gravity is compensated by means of frequency chirping between the Raman lasers, where we obtain the value of gravity as: $g=\frac{2\pi}{k_\text{eff}}\frac{df}{dt}$. By varying the chirping rate $\frac{df}{dt}$, we obtain interferometer fringe at $T=140$ ms, as shown in Figure~\ref{fig:fringe}. By sine fitting on the fringe at $T=140$ ms (with active vibration cancellation), we obtain an acceleration resolution of approximately $7.5$ \textmu{}Gal after 100 s (1 \textmu{}Gal = $10^{-8}$ m/s$^{-2}$ $\approx$ $10^{-9}$g), equivalent to a sensitivity of approximately $7.6\times 10^{-8}$ g/$\sqrt{\text{Hz}}$. Allan deviations of the gravity measurements (after subtracting the expected tidal variations) are performed by fixing the Raman frequency chirping rate at the steepest slope (Figure~\ref{fig:fringe}). Figure~\ref{fig:tidal} shows an uninterrupted record of local gravity variations for a period of more than 90 hours.

\par

\section{Conclusions}
We demonstrate a compact inertial stabilization system that provides vertical vibration cancellation as well as tilt stabilization in two axes for the retro-reflecting mirror employed in an atomic gravimeter. High-precision measurements are performed using the transportable atomic gravimeter, with resolution reaching $2.8\times 10^{-9}$g after 4000 s of integration time. In the near future, we plan to transport the atomic gravimeter out of the laboratory and carry out gravitational
field mapping of numerous geologically interesting places, such as around the hot spring
of Singapore, or detecting underground voids and sink holes. We also envision moving the gravimeter to
locations with active volcanoes for monitoring underground magma movements.

\begin{acknowledgments}
We thank Christoph Hufnagel, Nathan Shetell, Kaisheng Lee and Koon Siang Gan for their careful reading of our manuscript and their many insightful comments and suggestions. This work is supported by the National Research Foundation (Singapore) under the Quantum Engineering Program, DSO National Laboratories (Singapore) and the Ministry of Education of Singapore.
\end{acknowledgments}

\bibliography{biblio}

\end{document}